
\global\newcount\meqno
\def\eqn#1#2{\xdef#1{(\secsym\the\meqno)}
\global\advance\meqno by1$$#2\eqno#1$$}
%
\global\newcount\refno
\def\ref#1{\xdef#1{[\the\refno]}
\global\advance\refno by1#1}
\global\refno = 1
\vsize=7.5in
\hsize=5in
\magnification=1200
\tolerance 10000
%

%

\def\bx{\partial^2}
\def\bzk{\bar Z_k(\Phi)}

\def\calh{ {\cal H}}

\def\calo{{\cal O}}

\def\dss{\int_0^{\infty}{ds\over s}}
\def\dmu{D_{\mu}}
\def\dnu{D_{\nu}}
\def\evv{\Bigl(e^{-V''(\Phi)s}-e^{-V''(0)s}\Bigr)}

\def\half{{1\over 2}}
\def\hp{\hat p}
\def\hv{\hat V}

\def\kak{k{\partial\rho^{(d,0)}_k\over {\partial k}}}
\def\kbk{k{\partial\rho^{(d,1)}_k\over {\partial k}}}
\def\krk{k{\partial\rho^{(d)}_k(s,\Lambda)\over {\partial k}}}
\def\ksk{k{\partial{\tilde S_k[\Phi]}\over\partial k}}

\def\kuk{k{{\partial U_k(\Phi)}\over {\partial k}}}
\def\kzk{k{{\partial Z_k(\Phi)}\over\partial k}}

\def\mapright#1{\smash{\mathop{\longrightarrow}\limits^{#1}}}

\def\parmu{\partial_{\mu}}
\def\parnu{\partial_{\nu}}
\def\pmu{p_{\mu}}
\def\pnu{p_{\nu}}
\def\pv{( p^2+ V'')}
\def\pvh{(\hat p^2+\hat V'')}
\def\rhok{ \rho^{(d)}_k(s, \Lambda)}

\def\sumz{\sum_{n=0}^{\infty}}

\def\tsk{\tilde S_k[\Phi]}

\def\zk{ Z_k(\Phi)}
\def\pmb#1{\setbox0=\hbox{$#1$}%
\kern-.025em\copy0\kern-\wd0
\kern.05em\copy0\kern-\wd0
\kern-.025em\raise.0433em\box0 }
%
%
\baselineskip=0.1cm
\medskip
\nobreak
\medskip

\baselineskip 12pt plus 1pt minus 1pt
\vskip 2in
\centerline{\bf ON THE CONNECTION BETWEEN MOMENTUM CUTOFF}
\vskip 12pt
\centerline{{\bf AND OPERATOR CUTOFF REGULARIZATIONS}
\footnote{*}{This work is
supported in part by funds
provided by the U. S. Department of Energy (D.O.E.) under contract
\#DE-FG05-90ER40592.}}
\vskip 24pt
\centerline{{Sen-Ben Liao}
\footnote{$^{\dagger}$}{email: senben@phy.duke.edu}}
\vskip 12pt
\centerline{\it Department of Physics}
\centerline{\it Duke University }
\centerline{\it Durham, North Carolina\ \ 27708\ \ \ U.S.A.}
\vskip 12pt

\vskip 1.3in
\vskip 24 pt
\vskip 1.in
\baselineskip 12pt plus 2pt minus 2pt
\centerline{{\bf ABSTRACT}}
\medskip
\medskip
Operator cutoff regularization based on the original
Schwinger's proper-time formalism is examined.
By constructing
a regulating smearing function for the proper-time integration,
we show how this regularization scheme simulates the usual momentum cutoff
prescription yet preserves gauge symmetry even in the presence of the cutoff
scales.
Similarity between the operator cutoff regularization and the method of higher
(covariant) derivatives is also observed. The invariant
nature of the operator cutoff regularization makes it a promising tool
for exploring the renormalization group flow of gauge theories in the
spirit of Wilson-Kadanoff blocking transformation.

\vskip 24pt
\vfill
\noindent DUKE-TH-94-80\hfill October 1995
\eject

\centerline{\bf I. INTRODUCTION}
\medskip
\nobreak
\xdef\secsym{1.}\global\meqno = 1
\medskip
\nobreak

An essential step for identifying the physical contents of quantum field
theory is the removal of ultraviolet (UV) divergences which are due
to the presence
of interactions. The procedure, known as renormalization, operates on the
use of some regularization schemes to control the infinities followed by
a redefinition of the parameters contained in the original lagrangian in such
a way that physical quantities are independent of the regularization choice.
Although various methods are available,
it is often desirable to choose one which respects all the symmetry
properties present in the original theory. For
example, when studying gauge theories such as QCD or QED, a
momentum cutoff regulator would not be appropriate since it explicitly
violates gauge invariance. One must therefore turn to gauge invariant
prescriptions such as dimensional regularization \ref\thooft, $\zeta$
function regularization \ref\dowker, the proper-time method \ref\schwinger,
or the
Pauli-Villars procedure \ref\pauli. On the other hand, due to the
dimensionality dependence on the definition of $\gamma_5$, difficulties
are encountered when applying dimensional regularization to chiral theories.

Despite the shortcoming of its gauge non-invariant nature,
momentum cutoff has proven to be a useful regulator. Besides being simple and
more physical, one not only can immediately identify
the divergent structures of the theory, but also derive
readily the renormalization group (RG) flow equations which in turn
give predictions to how the theory behaves in different momentum regimes.
The renown RG formalism pioneered by Wilson and Kadanoff \ref\wilson\
is based on the use of
this regulator. In addition, when probing physics in the infrared (IR),
it is often advantageous to derive a low-energy effective
theory by integrating out the irrelevant short-distance modes.
The scale that separates the fast-fluctuating short-distance
modes from the slowly-varying components appears naturally in the
momentum cutoff regularization.

Momentum cutoff regularization can be formulated systematically
by means of blocking transformation \ref\lp\ - \ref\wetterich. To illustrate
the idea, consider the
scalar field theory as an example. From the original field $\phi(x)$
we first define a coarse-grained averaged blocked field $\phi_k(x)$
for each given
block of size $k^{-d}$ in $d$ dimensional Euclidean space with a
smearing function $\rho^{(d)}_k(x)$ as:
\eqn\pav{ \Phi(x)=\phi_k(x)=\int_y\rho^{(d)}_k(x-y)\phi(y),
\qquad \int_x=\int d^dx.}
The role of $\rho^{(d)}_k(x)$ is to provide an averaging of the fields within
the block and retain the degrees of freedom that are relevant for studying
the physics near the energy scale $\sim k$. Having defined
$\phi_k(x)$, the corresponding blocked action
$\tilde S_k[\Phi]$ can be written as \ref\finite
\eqn\cactt{ e^{-\tilde S_k[\Phi]}=\int D[\phi]\prod_x
\delta(\phi_k(x)-\Phi(x))e^{-S[\phi]}.}
In the above, the infinite product of $\delta$-functions strictly
 speaking only makes sense on the lattice where there are fewer blocked
 fields compared to the original fields. To render the procedure well defined,
 one may first replace this product by a Gaussian function \ref\ma
 \eqn\intewe{ \prod_x\delta(\phi_k(x)-\Phi(x))\to {\rm exp}\Bigl\{
 -\omega\int d^dx\bigl[\phi_k(x)-\Phi(x)\bigr]^2\Bigr\},}
 where $\omega$ is a large constant of dimension $({\rm mass})^2$.
 Physical limit corresponds to taking $\omega\to\infty$.
 It is also known that if $\rho^{(d)}_k(x)$ is
 a smooth Gaussian function,
 there will be no sharp boundary between the integrated and unintegrated
 modes \wilson. On the other hand,
 the desired scale of separation is naturally set at $p=k$ if one chooses
 \eqn\stfun{ \rho_k(x)=\int_{|p| < k}{d^dp\over(2\pi)^d}e^{-ipx},}
 or, $\rho_k(p)=\Theta(k-p)$, i.e. a sharp step function
 in momentum space. Although such a sharp cutoff will in general produce
 nonlocal interactions in $\tilde S_k[\Phi]$, the RG flow equation
 based on the infinitesimal variation of $k$ can still be formulated, and
 has been successfully carried out by Wegner and Houghton in
 \ref\wegner. When no
 confusion arises, we adopt the same general notation
 $\rho_k$ for both coordinate and momentum space representations,
 and distinguish between them by the arguments they carry.

 Choosing \stfun\ as the smearing function, the Fourier modes can be
 decomposed into
 \eqn\field{\phi(p)=\cases{\phi_{<}(p),&$0 \le p \le k$ \cr
 \cr
 \phi_{>}(p), &$k < p < \Lambda$, \cr }}
where $\phi_{<}$ and $\phi_{>}$ are, respectively, the slow and the
fast modes.
This in turn implies
\eqn\cacttm{\phi_k(p)=\rho_k(p)\phi(p)=\phi_{<}(p),}
which shows clearly how the fast modes are completely integrated
over through blocking transformation. Since it is generally a hopeless task to
compute $\tsk$ exactly, the complicated expression
in \cactt\ is frequently approximated by loop expansion. At the one-loop
level, we have
\eqn\efact{\eqalign{\tilde S_k[\Phi]&=-{\rm ln}\int D[\phi_{<}]
D[\phi_{>}]\prod_x\delta(\phi_k(x)-\Phi(x))~{\rm exp}\Bigl\{-S[\phi_{<}+
\phi_{>}]\Bigr\}\cr
&
=-{\rm ln}\int D[\phi_{<}]\prod_p\delta(\phi_{<}(p)-\Phi(p))\int D[\phi_{>}]\cr
&\qquad\qquad\qquad\times
{}~{\rm exp}\Bigl\{-S[\phi_{<}]-{1\over 2}\int_p^{'}\phi_{>}(p)K(\phi_{<})
\phi_{>}(-p)+\cdots\Bigr\} \cr
&
=-{\rm ln}\int D[\phi_{<}]\prod_p\delta(\phi_{<}(p)-\Phi(p))~{\rm exp}
\Bigl\{-S[\phi_{<}]-{1\over 2}{\rm Tr'}{\rm ln}K(\phi_{<})+\cdots\Bigr\} \cr
&
=S[\Phi]+~\half{\rm Tr'}{\rm ln}K(\Phi)+\cdots,}}
where
\eqn\kerdd{ K(\Phi) = {{\partial^2 S}
\over {\partial\phi(x)\partial\phi(y)}}\Big\vert_{\Phi}=
\Bigl(-\partial^2 +V''(\Phi)\Bigr)\delta^d(x-y),}
 \eqn\ssd{\int_p^{'}=\int_k^{\Lambda}{d^dp\over (2\pi)^d}
 =S_d\int_k^{\Lambda}dp~p^{d-1},\qquad S_d={2\over{(4\pi)^{d/2}\Gamma(d/2)}},}
 and ${\rm Tr'}$ implies taking the trace over a restricted momentum
 range $k \le p \le \Lambda$ as well as all possible internal indices.
 Without the prime
 notation, a complete momentum integration from zero to infinity is implied.
 Physically $\tilde S_k[\Phi]$ can be interpreted as an effective action
 parameterized by the averaged field $\Phi$ at the scale $k$, and it
 provides a smooth interpolation between the bare
 action $S[\Phi]$ defined at $k=\Lambda$ and the renormalized effective action
 $\tilde S_{k=0}[\Phi]$ which generates one-particle-irreducible
 Feynman graphs. Thus, the RG flow pattern of the theory is readily obtained
 by studying the change of $\tilde S_k[\Phi]$ in response to an infinitesimal
 change of the IR scale $k$.

 With the advantages of choosing the sharp cutoff regulator \stfun,
 one then inquires how it can be possible to extend this
 scheme to other theories possessing additional symmetries.
 Such a formulation will have profound implications on
 gauge theories such as QCD, QED, supersymmetry or quantum gravity.
 It may even offer new insights to the
 longstanding issue of quark confinement in the IR limit of strong
 interaction since the approach naturally yields an effective low-energy QCD
 lagrangian upon integrating out systematically the short-distance
 modes \ref\cwetterich. Non-perturbative effects can be explored, too.
 There will be new higher order interactions which are initially absent
 from the original lagrangian, and they may be of great import or
 even dominate in the IR regime despite the suppression
 at high energy.

Unfortunately, deriving the RG equation based on the
use of momentum cutoff regulator is known to conflict with gauge symmetry.
Yet the widely used dimensional regularization obscures the characteristics
of the Wilson-Kadanoff RG albeit it is an invariant prescription.
The first step toward applying the Wilson-Kadanoff RG to gauge theories is the
implementation of both the UV and the IR cutoff scales without destroying
gauge invariance. In \ref\michael, Oleszczuk demonstrated how this can be
achieved via blocking transformation in a completely symmetry-preserving
manner.
The methodology of the ``operator cutoff regularization'' elegantly
presented there
relies on the construction of a smooth smearing function $\rho(\Lambda^2 s)$,
where $\Lambda$ is to be identified with the usual UV regulator,
and $s$ the proper-time variable carrying dimension $({\rm mass})^{-2}$.
Embedding the smearing function into the $s$ integration,
one is lead to the following regularized parameterization:
\eqn\hktw{ {\rm Tr}\Bigl({\rm ln}{\cal H}-{\rm ln}{\cal H}_0\Bigr)
\Big\vert_{\rm oc}
= -\int_0^\infty{ds\over s}\rho(\Lambda^2 s)~{\rm Tr}\Bigl(e^{-{\cal H}s}
-e^{-{\cal H}_0s}\Bigr) ,}
where $\calh$ is an arbitrary fluctuation operator governing the
quadratic fluctuations of the fields and $\calh_0$
its corresponding limit of vanishing background field.
The subscript ``oc'' stands for operator cutoff.
For bosonic theories, $\calh$ is a positive
definite elliptic operator and $\calh^{-1}$ defines the propagator.
With a suitable choice of $\rho(\Lambda^2 s)$, the conventional cutoff
results may be recovered.

In the present work we follow closely the techniques outlined in \michael\
and generalize the operator cutoff regularization to arbitrary dimension $d$
using the smearing function $\rho^{(d)}_k(s,\Lambda)$, where
$k$ will be shown to play the role of an effective IR cutoff for the theory.
In this manner, any possible divergence
originating from momentum integration, whether of UV or IR nature,
will be turned into a singularity in $s$ and subsequently regulated by
$\rhok$. However, unlike the momentum cutoff approach,
operator cutoff regularization
is an invariant regularization since the proper-time variable $s$
is independent
of gauge transformation. Nonetheless, we emphasize that it is an
invariant prescription provided that no cutoff scales are imposed on the
momentum integral and the $s$ integration is left as the last step.
If the $s$ integral is carried out first, divergences generated
from taking the spacetime trace will manifest in the $p$ integration
and one may be forced to use some non-invariant regularization prescriptions.
Even though the smearing procedure is now acting on the proper-time
variable $s$, we retain the same general notation $\rho_k$ here since
the role of $\rhok$ in $s$ is similar to what $\rho_k(x)$ [cf. \stfun]
does in the coordinate space. However, due to the difference in their origin,
the functional forms of $\rhok$ and $\rho^{(d)}_k(x)$ are expected to be
rather different. It is important to keep in mind that form of $\rhok$
is not unique at all;
prescriptions such as the Pauli-Villars regulator and
dimensional regularization can all be shown to fall under the generalized
class of proper-time by a suitable definition of smearing function
\ref\chiral. Lucid discussions on the applications of proper-time
regularization can also be found in \ref\zuk.

What we shall demonstrate in this paper is that with a particular choice of
$\rho^{(d)}_k(s,\Lambda)$
operator cutoff regularization reproduces the usual one-loop blocked potential
$U_k(\Phi)$ which contains the IR cutoff scale $k$. When
considering $U_k(\Phi)$ in terms of Feynman diagrams, both
operator cutoff and momentum cutoff regularizations yield the
same results order by order in terms of coupling constant. The spirit of
our operator cutoff
formalism presented here is parallel to the idea of ``invariant
momentum space regularization'' treated by Ball in \chiral. However,
when the full blocked action is considered, deviation between momentum
cutoff and operator cutoff
prescriptions occur in the higher order (covariant) derivative terms. We find
that the effective blocked action regularized with the former contains gauge
noninvariant terms which are completely absent in the latter.

The organization of the paper is as follows: In Sec. II using scalar
theory as an example
we derive the generalized smearing function $\rho^{(d)}_k(s,\Lambda)$
which provides the bridge for establishing the functional equivalence
between the momentum cutoff and the operator cutoff
regularizations at the level of one-loop blocked potential.
Similarity between the operator cutoff and the Pauli-Villars
regularizations is discussed. Attempt to equate the two regularization
schemes beyond the leading order blocked potential is made in Sec. III.
It is found that at each order in the derivative expansion a new
smearing function must be introduced in the operator cutoff formalism
in order to give the same differential flow equations as that provided by the
sharp cutoff. In general, to ensure equality between the two
schemes to arbitrary order
${(\partial^2)}^n$, we need a total of $n+1$ smearing functions; i.e.
$\rho_k^{(d,m)}(s,\Lambda)$ with $m=0,1,\cdots, n$.
In Sec. IV we first reexamine the gauge
noninvariant nature of the momentum cutoff regularization by identifying
explicitly the symmetry violating components and their corresponding
proper-time counterparts. An invariant operator
cutoff scheme is then proposed to eliminate the gauge noninvariant sector
and restore the symmetry. Similarity between the operator cutoff
regularization and the method of higher covariant derivatives can also be
inferred. Section V is reserved for summary and discussions.

\medskip
\bigskip
\centerline{\bf II. OPERATOR CUTOFF REGULARIZATION }
\medskip
\nobreak
\xdef\secsym{2.}\global\meqno = 1
\medskip
\nobreak

Consider for simplicity the following one-component scalar lagrangian:
\eqn\scala{ {\cal L}={1\over 2}(\parmu\phi)^2+V(\phi).}
The one-loop contribution to the blocked action can be written as
\eqn\uer{ \tilde S_k^{(1)}[\Phi]={1\over 2}{\rm Tr'}{\rm ln}~K(\Phi)
={1\over 2}\int_x\int_p^{'}{\rm ln}
\Bigl({{p^2+V''(\Phi)}\over {p^2+V''(0)}}\Bigr),}
where $\Phi$ is the blocked field. Consider for simplicity
the low-energy limit where
the blocked action can be approximated by derivative expansion:
\eqn\lowsd{ \tilde S_k[\Phi]=\int_x\Biggl\{ {Z_k(\Phi)\over 2}(\parmu\Phi)^2
+U_k(\Phi)+O(\partial^4)\Biggr\},}
with $Z_k(\Phi)$ being the wavefunction renormalization constant.
The leading order contribution is then the one-loop blocked potential:
\eqn\ukone{ U_k^{(1)}(\Phi)={1\over 2}\int_p^{'}{\rm ln}
\Bigl({{p^2+V''(\Phi)}\over {p^2+V''(0)}}\Bigr).}
Differentiating the above with respect to the arbitrary IR scale $k$
leads to the following flow equation:
\eqn\renc{ k{{\partial U_k(\Phi)}\over{\partial k}}= -{S_d\over 2}k^d~
{\rm ln}\Bigl({{k^2+V''(\Phi)}\over{k^2+V''(0)}}\Bigr).}
This linear differential equation is obtained based on the so-called
``independent-mode approximation'' since it incorporates only the
contribution from
one-loop order and ignores the continuous feedbacks between different modes.
A RG improved equation which takes the interactions
between fast and slow
modes into consideration is given by the following modified expression
\ref\lpp:
\eqn\rennc{ k{{\partial U_k(\Phi)}\over{\partial k}}= -{S_d\over 2}k^d~
{\rm ln}\Bigl({{k^2+U_k''(\Phi)}\over{k^2+U_k''(0)}}\Bigr).}
In \ref\ft\ where field theory at finite temperature was considered,
it was found that
the independent mode approximation breaks down in the high temperature limit
and one must resort to the finite-temperature RG improved equation similar to
\rennc\ in order to account for the important daisy and superdaisy graphs.

However, when gauge theories are considered, momentum cutoff regulator
is not directly applicable for generating a flow equation such as
\rennc\ since it does not respect gauge symmetry.
The key issues which we wish
to explore are the following: Are cutoff scales {\it truly} in conflict
with gauge symmetry? Can we formulate a scheme
which contains cutoffs yet allows for the investigation of RG
flow for gauge theories in the spirit of Wilson-Kadanoff blocking
transformation? We now turn to the operator cutoff regularization
which offers the hope of introducing the cutoff scales in a
symmetry-preserving manner.

The basis of the operator cutoff formalism is provided by
Schwinger's proper-time regularization \schwinger\ in which one employs the
following identity for computing the one-loop contribution:
\eqn\hkorr{ {\rm Tr}\Bigl({\rm ln}{\cal H}-{\rm ln}{\cal H}_0\Bigr)
= -\int_0^\infty{ds\over s}{\rm Tr}\Bigl(e^{-{\cal H}s}
-e^{-{\cal H}_0s}\Bigr).}
The idea of operator cutoff regularization is to modify the above expression
by introducing into the proper-time integration a regulating smearing
function $\rhok$ such that
\eqn\hktw{ {\rm Tr'}\Bigl({\rm ln}{\cal H}-{\rm ln}{\cal H}_0\Bigr)
\longrightarrow {\rm Tr}\Bigl({\rm ln}{\cal H}-{\rm ln}{\cal H}_0\Bigr)
\Big\vert_{\rm oc}
= -\int_0^\infty{ds\over s}\rhok~{\rm Tr}\Bigl(e^{-{\cal H}s}
-e^{-{\cal H}_0s}\Bigr) ,}
i.e., a complete trace can now be taken after inserting
$\rhok$ into the $s$ integration.
The absence of any cutoff in the $p$ integration is a {\it sine qua non}
for preserving gauge symmetry.
As an illustration, we consider the lagrangian in \scala.
Following the procedures outlined in \michael\ and using $\calh=p^2+V''(\Phi)$,
the one-loop blocked potential becomes:
\eqn\ukky{\eqalign{ U_k^{(1)}(\Phi)&={1\over 2}\int_p^{'}~
{\rm ln}\Bigl({{p^2+V''(\Phi)}\over {p^2+V''(0)}}\Bigr) \cr
&
\longrightarrow -{1\over 2}\dss\rhok\int_pe^{-p^2s}\evv \cr
&
=-{1\over 2(4\pi)^{d/2}}\int_0^{\infty}{ds\over s^{1+d/2}}\rhok\evv,}}
where we have used
\eqn\spint{ \int_p {(p^2)}^me^{-p^2s}
={\Gamma(m+d/2)\over {(4\pi)^{d/2}\Gamma(d/2)}}s^{-(m+d/2)}.}
The cutoffs are now taken over by the smearing function.
With the $k$ dependence of $\tsk$ contained entirely in $\rhok$, probing the
RG flow of the theory amounts to studying the change of $\rhok$ with
varying the IR cutoff $k$. Thus, the differential flow equation of the
theory can be written as:
\eqn\fll{ k{{\partial U_k(\Phi)}\over{\partial k}}=
-{1\over 2(4\pi)^{d/2}}\int_0^{\infty}{ds\over s^{1+d/2}}
\bigl(\krk\bigr)\evv.}
To deduce the form of $\rhok$, we now equate \fll\ with \renc\ which
is derived using the cutoff approach:
\eqn\flo{\eqalign{\kuk &=-{S_d\over 2} k^d~
{\rm ln}\Bigl({{k^2+V''(\Phi)}\over{k^2+V''(0)}}\Bigr)
={S_d\over 2}k^d\dss e^{-k^2s}\evv \cr
&
=-{1\over 2(4\pi)^{d/2}}\int_0^{\infty}
{ds\over s^{1+d/2}}\bigl(\krk\bigr)\evv ,}}
or equivalently,
\eqn\flr{ \krk=-{2\over\Gamma(d/2)}(k^2s)^{d/2}e^{-k^2s}.}
We shall choose
a set of boundary conditions for $\rho^{(d)}_k(s,\Lambda)$ which
renders \ukky\ finite throughout the calculation.
Since the proper-time variable $s$ has dimension $({\rm length})^2$,
UV divergence corresponding to the short-distance singularity appears at $s=0$.
Thus, setting $\rho_k^{(d)}(s=0,\Lambda)=0$ will eliminate the unwanted
UV singularity as $s\to 0$. On the other hand, since we wish to
modify only the UV
behavior of the theory while leaving the IR physics intact, it is appropriate
to have $\rho_{k=0}^{(d)}(s\to\infty,\Lambda)=1$. Finally, we demand
$\rho_{k=\Lambda}^{(d)}(s,\Lambda)=0$ since the one-loop contribution must
vanish at the UV cutoff scale $\Lambda$ to give back the
original bare
theory. Solving \flr\ subject to the conditions imposed above leads to:
\eqn\rro{\eqalign{ \rhok &=\rho^{(d)}(\Lambda^2s)-\rho^{(d)}(k^2s)
={2s^{d/2}\over\Gamma(d/2)}\int_k^{\Lambda}dz~ z^{d-1}
e^{-z^2s}={2s^{d/2}\over{S_d\Gamma(d/2)}}\int_z^{'}e^{-z^2s} \cr
&
={1\over\Gamma(d/2)}\Gamma[{d\over 2};k^2s,\Lambda^2s],}}
where
\eqn\gam{ \Gamma[\alpha;x_1,x_2]=\int_{x_1}^{x_2}dx~x^{\alpha-1}e^{-x}}
is the generalized incomplete $\Gamma$ function.
Notice that in the physical limit $k\to 0$ and $\Lambda\to\infty$, we have
\eqn\smew{ \rho_{k=0}^{(d)}(s,\Lambda\to\infty)=1,}
and the operator cutoff regularization is reduced to that
of Schwinger's
proper-time. In this limit
UV and possible IR divergences may appear and additional counterterms
must be added to subtract off the infinities \schwinger.

To explicitly demonstrate that our smearing function $\rhok$ simulates
a sharp momentum cutoff regulator, we substitute \rro\
into the last expression of \ukky:
\eqn\ukt{U_k^{(1)}(\Phi) =- {1\over 2}
\dss\evv\int_z^{'}e^{-z^2s} ,}
which, upon switching the order of integrations between $s$ and $z$
and equating $z$ with the momentum variable $p$, gives back \uer.
Thus, we conclude that the proper-time smearing function derived in \rro\
completely reproduces the usual blocked potential $U_k(\Phi)$ at the one-loop
level. That the dummy integration variable
$z$ hidden in $\rhok$ turns out to be the momentum variable $p$
can be seen from a direct substitution of
\rro\ into the second equation on the right-hand-side of \ukky\ which yields
\eqn\uke{\eqalign{U_k^{(1)}(\Phi) &=-{1\over{S_d\Gamma(d/2)}}
\int_x\int_0^{\infty}{ds\over s^{1-d/2}}\evv \cr
&\qquad\qquad
\times \int_z^{'}e^{-z^2s}\int_pe^{-p^2s}.}}
The equation readily shows how $p$ is intimately connected to $z$ through
the transfer of cutoff dependence.

It is instructive to examine how the propagators
and the one-loop kernel are modified in the presence of $\rhok$.
Straightforward calculation leads to
\eqn\hkk{\eqalign{ {1\over \calh^n}&={1\over\Gamma(n)}\int_0^\infty ds~
s^{n-1}e^{-{\cal H}s}~\mapright{{\rm oc}}~
{1\over \calh^n}\Big\vert_{\rm oc}={1\over\Gamma(n)}
\int_0^\infty ds~s^{n-1}\rho_k^{(d)}(s,\Lambda)e^{-{\cal H}s} \cr
&
={1\over\calh^n}\cdot{{2\Gamma(n+d/2)}\over {d\Gamma(n)\Gamma(d/2)}}
\Biggl\{\Bigl({\Lambda^2\over\calh}\Bigr)^{d/2}F\Bigl({d\over 2},
{d\over 2}+n,1+{d\over 2};-{\Lambda^2\over\calh}\Bigr) \cr
&\qquad\qquad\qquad\qquad~~
-\Bigl({k^2\over\calh}\Bigr)^{d/2}F\Bigl({d\over 2},
{d\over 2}+n,1+{d\over 2};-{k^2\over\calh}\Bigr)\Biggr\},}}
and
\eqn\hkorr{\eqalign{{\rm Tr'}\Bigl({\rm ln}{\cal H}-&{\rm ln}{\cal H}_0\Bigr)
= -\int_0^\infty{ds\over s}{\rm Tr}\Bigl(e^{-{\cal H}s}
-e^{-{\cal H}_0s}\Bigr) \cr
&
\longrightarrow {\rm Tr}\Bigl({\rm ln}{\cal H}-{\rm ln}{\cal H}_0\Bigr)
\Big\vert_{\rm oc}=-\int_0^\infty{ds\over s}\rhok{\rm Tr}\Bigl(e^{-{\cal H}s}
-e^{-{\cal H}_0s}\Bigr) \cr
&
=-{2\over d}{\rm Tr}\Biggl\{\Bigl({\Lambda^2\over\calh}\Bigr)^{d/2}F\Bigl(
{d\over 2},
{d\over 2},1+{d\over 2};-{\Lambda^2\over\calh}\Bigr)
-\Bigl({\Lambda^2\over\calh_0}\Bigr)^{d/2}F\Bigl({d\over 2},
{d\over 2},1+{d\over 2};-{\Lambda^2\over\calh_0}\Bigr) \cr
&\qquad~~
-\Bigl({k^2\over\calh}\Bigr)^{d/2}F\Bigl({d\over 2},
{d\over 2},1+{d\over 2};-{k^2\over\calh}\Bigr)
+\Bigl({k^2\over\calh_0}\Bigr)^{d/2}F\Bigl({d\over 2},
{d\over 2},1+{d\over 2};-{k^2\over\calh_0}\Bigr)\Biggr\},}}
where
\eqn\funn{ F\Bigl(a,b,c;\beta\Bigr)=B^{-1}(b,c-b)\int_0^1 dx~x^{b-1}
(1-x)^{c-b-1}(1-\beta x)^{-a}}
is the hypergeometric function symmetric under the exchange between $a$
and $b$, and
\eqn\ber{ B(x,y)={\Gamma(x)\Gamma(y)\over\Gamma(x+y)}=\int_0^1 dt~t^{x-1}
(1-t)^{y-1},}
the Euler $\beta$ function. For $n=1$ and $2$, \hkk\ yields, respectively,
\eqn\hkko{ {1\over \calh}\Big\vert_{\rm oc}=
\int_0^\infty ds~\rho_k^{(d)}(s,\Lambda)e^{-{\cal H}s}
={1\over\calh}\Biggl\{\Bigl({\Lambda^2\over
{\calh+\Lambda^2}}\Bigr)^{d/2}-\Bigl({k^2\over{\calh+k^2}}\Bigr)^{d/2}
\Biggr\},}
and
\eqn\hkkoy{\eqalign{ {1\over \calh^2}\Big\vert_{\rm oc}&=
\int_0^\infty ds~s\rho_k^{(d)}(s,\Lambda)e^{-{\cal H}s} \cr
&
={1\over\calh^2}\Biggl\{\Bigl({\Lambda^2\over
{\calh+\Lambda^2}}\Bigr)^{d/2}\Bigl[1+{d\over 2}{\calh\over{\calh+\Lambda^2}}
\Bigr]-\Bigl({k^2\over{\calh+k^2}}\Bigr)^{d/2}\Bigl[1+{d\over 2}{\calh\over
{\calh+k^2}}\Bigr]\Biggr\}.}}
By further restricting ourselves to $d=4$ where
\eqn\rty{ \rho_k^{(4)}(s,\Lambda)=(1+k^2s)e^{-k^2s}-(1+\Lambda^2s)
e^{-\Lambda^2s},}
the operator cutoff regularized propagator and the one-loop kernel
take on the following structures:
\eqn\cher{{1\over \calh}\Big\vert_{\rm oc}={1\over{\calh+k^2}}-{1\over{\calh
+\Lambda^2}}-{\Lambda^2\over{(\calh+\Lambda^2)^2}}+{k^2\over {(\calh+k^2)^2}},}
and
\eqn\hko{\eqalign{ {\rm Tr}\Bigl({\rm ln}{\cal H}-{\rm ln}{\cal H}_0\Bigr)
\Big\vert_{\rm oc}&= {\rm Tr}\Biggl\{{\rm ln}\Bigl[ { {{\cal H}+k^2}\over
{{\cal H}_0+k^2}}\times
{{\calh_0+\Lambda^2}\over {\calh+\Lambda^2}}\Bigr]
-{\Lambda^2(\calh-\calh_0)\over {(\calh+\Lambda^2)
(\calh_0+\Lambda^2)}} \cr
&\qquad~~
+{k^2(\calh-\calh_0)\over {(\calh+k^2)(\calh_0+k^2)}}\Biggr\} .}}
The above equations imply that one may regard $\Lambda$ as the mass of
some unitarity-violating ghost states, which
can be seen from the relative negative sign in the
modified propagator. However, in the limit $\Lambda\to\infty$, the ghosts
decouple from the theory, as they should. The IR cutoff
scale $k$ which can also be thought of as being the ``fictitious'' mass
ascribed to the fields is a useful regulator particularly when the theory
contains massless
particles. Physical observables must be computed, however, by taking the
limit $k\to 0$. From \cher, one may also say that the effect of $\Lambda$
is equivalent to introducing higher derivative terms into the theory.
In other words,
the lagrangian density \scala\ may be replaced by the corresponding
regularized counterpart:
\eqn\counp{ {\cal L}_{\rm reg.}={1\over 2}\phi\Bigl[
-\bx+{2\over\Lambda^2}\bigl(-\bx\bigr)^2+{1\over\Lambda^4}
\bigl({-\bx}\bigr)^3\Bigr]\phi+V(\phi).}

The interpretations on the role played by the cutoff
scales in the operator cutoff approach are reminiscent to that
of the Pauli-Villars regulator. In fact, one can show that
the conventional Pauli-Villars scheme is a special case
of the proper-time regularization having a smearing function of the form \zuk:
\eqn\pvsmear{ \rho_k^{\rm pv}(s,\Lambda)=\sum_i\Bigl(a_ie^{-k_i^2s}
-b_ie^{-\Lambda_i^2s}\Bigr),}
with $\Lambda_i$ and $k_i$ carrying the same meaning as the operator cutoff
scales. In order to render the theory finite, the
coefficients $a_i$ and $b_i$ as well as $i$, the number
of ghost terms, must be appropriately chosen.
Eq. \pvsmear\ implies
\eqn\chpv{{1\over \calh^n}\Big\vert_{\rm pv}=\sum_i{1\over\Gamma(n)}
\int_0^{\infty}ds~s^{n-1}\Bigl(a_ie^{-k_i^2s}-b_ie^{-\Lambda_i^2s}\Bigr)
e^{-\calh s}=\sum_i\Bigl[{a_i\over{(\calh+k_i^2)^n}}-{b_i\over{(\calh
+\Lambda_i^2)^n}}\Bigr],}
and
\eqn\hpv{\eqalign{{\rm Tr}{\rm ln}\Bigl({\cal H}-{\cal H}_0\Bigr)
\Big\vert_{\rm pv}&=-\sum_i\int_0^{\infty}{ds\over s}\Bigl(a_ie^{-k_i^2s}
-b_ie^{-\Lambda_i^2s}\Bigr){\rm Tr}\Bigl(e^{-\calh s}-e^{-\calh_0 s}\Bigr) \cr
&
={\rm Tr}\sum_i{\rm ln}\Biggl[\Bigl({{{\cal H}+k_i^2}\over{{\cal H}_0
+k_i^2}}\Bigr)^{a_i}\times \Bigl(
{{\calh_0+\Lambda_i^2}\over {\calh+\Lambda_i^2}}\Bigr)^{b_i}\Biggr],}}
which resemble \cher\ and \hko\ by simply taking $a_i=b_i=i=1$. The
minute difference can be attributed to the
extra terms that are linearly dependent in $s$ in the definition of the
smearing function.

 For a general fluctuation operator of the form
 $\calh=\calh_0+\lambda\delta\calh$, one can easily show that
 $\rhok$ reproduces the usual momentum cutoff results
 order by order in $\lambda$ by comparing the general expansion
 \eqn\etrf{ {\rm Tr'}~\Bigl({\rm ln}\calh-{\rm ln}\calh_0\Bigr)
 ={\rm Tr'}~{\rm ln}\Bigl[{1+\lambda\calh_0^{-1}\delta\calh}\Bigr]
 =-\sum_{n=1}^{\infty}{(-\lambda)^n\over n}{\rm Tr'}\Bigl\{\bigl(\calh_0^{-1}
 \delta\calh\bigr)^n\Bigr\}}
with
\eqn\opcu{\eqalign{ {\rm Tr}~\Bigl({\rm ln}\calh-{\rm ln}\calh_0\Bigr)
\Big\vert_{\rm oc}&=-\sum_{n=1}^{\infty}{(-\lambda)^n\over n}
{\rm Tr}\Bigl\{\bigl({\calh_0^{-1}
\delta\calh}\bigr)^n\Bigr\vert_{\rm oc}\Bigr\} \cr
&
=-\int_0^{\infty}{ds\over s}\rhok {\rm Tr}\Bigl\{e^{-(\calh_0
+\lambda\delta\calh)s}-e^{-\calh_0s}\Bigr\} \cr
&
=-\int_0^{\infty}{ds\over s}\rhok~{\rm Tr}\Biggl\{e^{-\calh_0s}\Bigl[
\sum_{n=0}^{\infty}{(-\lambda)^n\over n!}\bigl(\delta\calh\bigr)^ns^n\Bigr]
-e^{-\calh_0s}\Biggr\} \cr
&
=-\sum_{n=1}^{\infty}{(-\lambda)^n\over n!}\int_0^{\infty}
dss^{n-1}\rhok~{\rm Tr}\Bigl\{e^{-\calh_0s}\bigl(\delta\calh\bigr)^n
\Bigr\}\cr
&
=-\sum_{n=1}^{\infty}{(-\lambda)^n\over n!}\int_0^{\infty}
dss^{n-1}{\rm Tr'}\Bigl\{e^{-\calh_0s}\bigl(\delta\calh\bigr)^n
\Bigr\}\cr
&
=-\sum_{n=1}^{\infty}{(-\lambda)^n\over n}{\rm Tr'}\Bigl\{\bigl(\calh_0^{-1}
\delta\calh\bigr)^n\Bigr\}}}
in the operator cutoff approach.
As an explicit check, we again consider the $\lambda\Phi^4$ theory
in $d=4$ with $k=0$ as an example and obtain the following one-loop
corrections to the two- and four-point vertex functions:
\eqn\contr{\eqalign{\delta\Gamma^{(2)}_{\rm oc} &=
{\lambda\over 2}\int_p{1\over{p^2+\mu^2}}
\Big\vert_{\rm oc}={\lambda\over 2}\int_p
{1\over{p^2+\mu^2}}\Bigl({\Lambda^2\over{p^2+\mu^2+\Lambda^2}}\Bigr)^2 \cr
&
={\lambda\over 32\pi^2}\Bigl[\Lambda^2-\mu^2{\rm ln}\Bigl({{\Lambda^2
+\mu^2}\over\mu^2}\Bigr)\Bigr]={\lambda\over 2}\int^{'}_p
{1\over{p^2+\mu^2}},}}
and
\eqn\contry{\eqalign{\delta\Gamma^{(4)}_{\rm oc}&=-{3\lambda^2\over 2}
\int_p{1\over{(p^2+\mu^2)}^2}\Big\vert_{\rm oc}=-{3\lambda^2\over 2}
\int_p{1\over{(p^2+\mu^2)}^2}\bigl({\Lambda^2\over{p^2+\mu^2+\Lambda^2}}
\bigr)^2\bigl[1+{2(p^2+\mu^2)\over{p^2+\mu^2+\Lambda^2}}\bigr]\cr
&
=-{3\lambda^2\over 32\pi^2}\Bigl[{\rm ln}\Bigl({{\Lambda^2+\mu^2}
\over\mu^2}\Bigr)-{\Lambda^2\over{\mu^2+\Lambda^2}}\Bigr]
=-{3\lambda^2\over 2}\int_p^{'}{1\over{(p^2+\mu^2)}^2},}}
which are in complete agreement with the momentum cutoff results.
Thus, we see that operator cutoff regularization reproduces the usual
momentum cutoff results order by order in the coupling constant expansion.

\medskip
\bigskip

\centerline{\bf III. OPERATOR CUTOFF AND DERIVATIVE EXPANSION }
\medskip
\nobreak
\xdef\secsym{3.}\global\meqno = 1
\medskip
\nobreak

It was demonstrated in the last Section that the smearing function $\rhok$
derived in \rro\ imitates the momentum cutoff at
the level of one-loop blocked potential $U_k(\Phi)$. In this Section we take
into account the $\partial^2$ derivative term and inquire how
to further construct the linkage between the two regularization schemes.
For the computation of the wavefunction
renormalization constant $Z_k(\Phi)$,
a small inhomogeneity is assumed to be present in the background
field. Various methods for deriving $\zk$ are available \ref\grad.
Below we rederive $\zk$ using the two prescriptions prescribed above
and compare their results. As we shall see, discrepancy between the
two schemes appears already at the level of $Z_k(\Phi)$
if only one smearing function $\rhok$ is used throughout.
Equality between the two formalisms up to
$O(\partial^2)$ is restored provided that an additional
smearing function be used.

\medskip
\centerline{\bf a. momentum cutoff regularization}
\medskip

To compute $Z_k(\Phi)$ via the momentum cutoff regularization, we
adopt the approach originated by Fraser \ref\fraser. The manner in which
the derivative terms are extracted is based on the notion of
treating the momentum
and field variables as non-commuting operators $\hat p_{\mu}$ and $\hat\Phi$
obeying the commutation relations:
\eqn\commut{[~ \hat\Phi,~\hat p_{\mu}~]~ = -i\parmu\hat\Phi ,}
and
\eqn\comu{ [~\hat\Phi,~\hat p^2~]~= -\bx \hat\Phi
- 2i\hat p_{\mu}\parmu \hat\Phi,}
where a caret symbol has been added to the operators to distinguish them
from the ordinary $c$-number variables. Repeated use of the above
relations leads to
\eqn\esr{\eqalign{ (\hat p^2+\hat V'')^n &=:\pvh^n:-{1\over 2}n(n-1):
\pvh^{n-2}:\bx \hat V'' \cr
&
-in(n-1)\hat\pmu:\pvh^{n-2}:\parmu \hat V'' \cr
&
-{1\over 3}n(n-1)(n-2):\pvh^{n-3}:\bigl(\parmu \hat V''\bigr)^2 \cr
&
-{2\over 3}n(n-1)(n-2)\hat\pmu\hat\pnu :\pvh^{n-3}:\parmu\parnu \hat V''  \cr
&
-{1\over 2}n(n-1)(n-2)(n-3)\hat\pmu\hat\pnu :\pvh^{n-4}:\parmu \hat V''
\parnu \hat V'' +O(\partial^4),}}
where $:~:$ implies a ``normal ordering'' procedure such that
all $\hat p$ dependences are moved to the left of the $\hat\Phi$-dependent
terms.
This is a necessary step for evaluating the functional trace in \uer\
since the momentum integration is to be performed before $x$. Once the normal
ordering procedure is done, we may simply drop the carets and
treat the quantities on the right-hand-side of \esr\ as
ordinary $c$-numbers since any further application of \commut\ or \comu\
will only generate higher order derivative terms which do not affect the
computation of $Z_k$.

With the help of the identity
\eqn\erte{ {\rm ln}\bigl( \hat p^2+ \hat V''\bigr)= \lim_{n\to 0}
{\partial\over\partial n}\bigl( \hat p^2+\hat V''\bigr)^n,}
the one-loop contribution to the blocked action becomes (dropping the carets)
\eqn\effe{\eqalign{ \tilde S^{(1)}_k[\Phi]&={1\over 2}\int_x\int_p^{'}\Biggl\{
{\rm ln}\pv+{{(9-2d)p^2-2dV''}\over{3d\pv^4}}(\parmu V'')^2 \cr
&\qquad\qquad
+{{(3d-8)p^2+3dV''}\over{6d\pv^3}}\bx V'' +O(\partial^4)\Biggr\} \cr
&
={1\over 2}\int_x\int_p^{'}\Biggl\{
{\rm ln}\pv+{{(d-3)p^2+dV''}\over{3d\pv^4}}(\parmu V'')^2+O(\partial^4)
\Biggr\}.}}
The above expression is obtained by first dropping the surface terms with
\eqn\vann{ 0=\int_x~\bx\Bigl[{1\over\pv^n}\Bigr]=\int_x\Biggl\{{n(n+1)\over
{\pv^{n+2}}}(\parmu V'')^2-{n\over\pv^{n+1}}\bx V''\Biggr\},}
followed by simplifying the momentum integrations using the the $O(d)$
invariant property:
\eqn\ods{\int_p^{'}p_{\mu_1}p_{\mu_2}\cdots p_{\mu_{2m}}f(p^2)=
{{T^m_{\mu_1\mu_2\cdots\mu_{2m}}\Gamma(d/2)}\over{2^m\Gamma(m+d/2)}}
\int_p^{'}{(p^2)}^m f(p^2),}
where $f(p^2)$ is an arbitrary scalar function and
\eqn\npoer{ T^m_{\mu_1\mu_2\cdots\mu_{2m}}=\delta_{\mu_1,\mu_2}\cdots
\delta_{\mu_{2m-1},\mu_{2m}}+ {\rm~ permutations}.}
While the first term in the last equation of \effe\ matches the
usual one-loop logarithmic
contribution for the blocked potential $U_k(\Phi)$, the second term
represents the correction to $\zk$. Taking the familiar $\lambda\Phi^4$ theory
as an example, with $V''(\Phi)=\mu^2+\lambda\Phi^2/2$, we have
\eqn\zss{ Z_k^{(1)}(\Phi)={\lambda^2\Phi^2\over 3d}\int_p^{'}
{(d-3)p^2+dV''\over\pv^4} ,}
which, upon differentiating with respect to $k$, yields the following
differential flow equation:
\eqn\renz{ \kzk = -{S_d\over 3d}k^d~\lambda^2\Phi^2{{(d-3)k^2+dV''}\over
{\bigl(k^2+V''\bigr)^4}}.}
For $d=4$, \zss\  becomes
\eqn\zza{ Z_k^{(1)}(\Phi)={\lambda^2\Phi^2\over 192\pi^2}
{{k^4+3k^2V''+{V''}^2}\over{\bigl(k^2+V''\bigr)^3}},}
which agrees with that obtained in \michael, \grad\ and \fraser\ in
the limit $k\to 0$.
Since the contribution to $\zk$ is UV finite, one can safely take
the limit $\Lambda\to\infty$.
\medskip
\medskip
\centerline{\bf b. operator cutoff regularization}
\medskip

In the alternative operator cutoff approach,
derivative terms also arise from commuting the operators $\hp_{\mu}$
and $\hat\Phi$.
The one-loop contribution to the blocked action in the proper-time
representation is given by:
\eqn\proo{ \tilde S^{(1)}_k[\hat\Phi]=-{1\over 2}\dss ~{\rm Tr'}\Bigl(
e^{-({\hat p}^2+{\hat V}''(\hat\Phi))s}-e^{-({\hat p}^2+{\hat V}''(0))s}
\Bigr),}
where the normal ordering procedure can be carried out as
\eqn\norw{\eqalign{ e^{-(\hp^2+\hv'')s}&=\sumz{{(-1)^n}\over n!}s^n
\pvh^n \cr
&
=~:e^{-(\hp^2+\hv'')s}:-:e^{-(\hp^2+\hv'')s}:
\times\Biggl\{ is^2\pmu\parmu \hv''+{1\over 2}s^2\bx \hv''
-{1\over 3}s^3\bigl(\parmu \hat V''\bigr)^2 \cr
&
-{2\over 3}s^3\hat\pmu\hat\pnu\parmu\parnu\hat V''
+{1\over 2}s^4\hat\pmu\hat\pnu\parmu \hat V''\parnu\hat V''+\cdots\Biggr\}\cr
&
\buildrel {O(d)}\over\longrightarrow~:e^{-(\hp^2+\hv'')s}:
-:e^{-(\hp^2+\hv'')s}:
\times\Biggl\{ {1\over 2}s^2\bx \hat V''
-{1\over 3}s^3\bigl(\parmu\hat V''\bigr)^2 \cr
&
-\hp^2\Bigl[{2s^3\over 3d}\bx\hat V''-{s^4\over 2d}\bigl(\parmu\hat V''\bigr)^2
\Bigr]+\cdots\Biggr\}.}}
Dropping the distinction between operators and $c$-numbers as before,
the one-loop correction to the blocked action becomes:
\eqn\trer{\eqalign{ \tilde S^{(1)}_k[\Phi]&=-{1\over2}
\int_x\int_0^{\infty}{ds\over s}
\int_p^{'} e^{-p^2s}\Bigl(e^{-V''s}-1\Bigr)\cr
&
+{1\over 12}\int_x\int_0^{\infty}ds~s~
\int_p^{'} e^{-p^2s}e^{-V''s}
\Bigl[3\bx V''-2s\bigl(\parmu V''\bigr)^2\Bigr]\cr
&
-{1\over 12d}\int_x\int_0^{\infty}ds~s^2
\int_p^{'}e^{-p^2s}p^2 ~e^{-V''s}\Bigl[
4\bx V''-3s\bigl(\parmu V''\bigr)^2\Bigr] .}}
In the above, the second and the third integrals together contribute to
$Z^{(1)}_k(\Phi)$. Eq.\trer\ reduces to \effe\ if the $s$
integration is carried out first and simplified by the help of \vann.
However, our goal here is to find out how the regulating smearing function(s)
should be implemented in the $s$ integrations in order to allow for
a complete $p$ integration without imposing any cutoff scales.
The caution to be taken here is that due to the different powers
of $p$ dependence, there is no reason {\it a priori} that the same smearing
function $\rhok$ can yield a wavefunction renormalization
constant $\zk$ identical to \renz\ which was derived from momentum cutoff.
That a new smearing function must be called for at each level of
derivative expansion is actually hinted from the following
integral transformation:
\eqn\uwed{\eqalign{ \int_p^{'}{{(p^2)}^m\over(p^2+a)^n}&={1\over\Gamma(n)}
\int_0^{\infty}ds~s^{n-1}e^{-as}\int_p^{'}{(p^2)}^m e^{-p^2s} \cr
&
\longrightarrow {1\over\Gamma(n)}\int_0^{\infty}ds~s^{n-1}e^{-as}
\rho_k^{(d,m)}(s,\Lambda)\int_p {(p^2)}^m e^{-p^2s} \cr
&
={\Gamma(m+d/2)\over{(4\pi)^{d/2}\Gamma(d/2)\Gamma(n)}}\int_0^{\infty}ds~
s^{n-1-m-d/2}e^{-as}\rho_k^{(d,m)}(s,\Lambda),}}
which is satisfied provided that
\eqn\tyre{\rho_k^{(d,m)}(s,\Lambda)
={2s^{m+d/2}\over{S_d\Gamma(m+d/2)}}\int_z^{'}{(z^2)}^me^{-z^2s}
={1\over{\Gamma(m+d/2)}}\Gamma[m+{d\over 2};k^2s,\Lambda^2s].}
One may explicitly check that by using $\rho_k^{(d,1)}(s,\Lambda)$ (with
$\rho_k^{(d,0)}(s,\Lambda)=\rho_k^{(d)}(s,\Lambda)$) for the third
integral in \trer, the original cutoff expression is recovered. With
\eqn\trert{\eqalign{ \tilde S^{(1)}_k[\Phi]&=-{1\over2}\int_x\int_0^{\infty}
{ds\over s}\rho_k^{(d,0)}(s,\Lambda)\int_p e^{-p^2s}\Bigl(e^{-V''s}
-1\Bigr) \cr
&
+{1\over 12}\int_x\int_0^{\infty}ds~s~\rho_k^{(d,0)}(s,\Lambda)
{}~\int_p e^{-p^2s}e^{-V''s}\Bigl[3\bx V''-2s \bigl(\parmu V''\bigr)^2\Bigr]
\cr
&
-{1\over 12d}\int_x\int_0^{\infty}ds~s^2\rho_k^{(d,1)}(s,\Lambda)~
\int_pe^{-p^2s}p^2~e^{-V''s}\Bigl[
4\bx V''-3s\bigl(\parmu V''\bigr)^2\Bigr],}}
upon performing the $p$ integrations followed by a differentiation with
respect to $k$, we arrive at
\eqn\fllo{\eqalign{ \ksk &= \int_x\biggl\{ \kuk+{1\over 2}\kzk
(\parmu\Phi)^2+\cdots\biggr\} \cr
&
=-{1\over{2(4\pi)^{d/2}}}
\int_x\int_0^{\infty}{ds\over s^{1+d/2}}\kak\Bigl(e^{-V''s}-1\Bigr) \cr
&
+{1\over{24(4\pi)^{d/2}}}
\int_x\bigl(\parmu V''\bigr)^2\int_0^{\infty}ds s^{2-d/2}~e^{-V''s}
\Bigl[2\kak -\kbk\Bigr],}}
where the expression is simplified by the help of
\eqn\nerr{ 0=\int_x \parmu\Bigl( e^{-V''s}\parmu V''\Bigr)
=\int_x e^{-V''s}\Bigl[\bx V''-\bigl(\parmu V''\bigr)^2s\Bigr].}
{}From \fllo, it is clear that despite the presence of the derivative terms
$\rho_k^{(d,0)}(s,\Lambda)$ is still given by \rro\ for the equality in
$k({\partial U_k/{\partial k}})$ to hold. As for $\zk$, comparison between
\fllo\ and \renz\ gives
\eqn\tztz{\eqalign{ \kzk &= -{S_d\over 3d}k^d\lambda^2\Phi^2{{(d-3)k^2+dV''}
\over{\bigl(k^2+V''\bigr)^4}} \cr
&={\lambda^2\Phi^2\over{12(4\pi)^{d/2}}}
\int_0^{\infty}ds s^{2-d/2}~e^{-V''s}\Bigl[2\kak-\kbk\Bigr].}}
Use of the identity
\eqn\idfg{ {1\over \calh^n}={1\over\Gamma(n)}\int_0^\infty ds~
s^{n-1}e^{-{\cal H}s}}
then leads to
\eqn\derf{\eqalign{\kbk&=2\kak+{4(k^2s)^{d/2}\over d\Gamma(d/2)}
\bigl(d-k^2s\bigr)e^{-k^2s} \cr
&
=-{4(k^2s)^{1+d/2}\over d\Gamma(d/2)}e^{-k^2s},}}
or
\eqn\rrgh{\rho_k^{(d,1)}(s,\Lambda)
={2s^{1+d/2}\over{S_d\Gamma(1+d/2)}}\int_z^{'}z^2e^{-z^2s}
={1\over{\Gamma(1+d/2)}}\Gamma[1+{d\over 2};k^2s,\Lambda^2s],}
which agrees with \uwed\ for $m=1$ and confirms that more than one
smearing function must be used to yield the same $Z_k(\Phi)$ in both
formalisms. Had $\rho_k^{(4)}(s,\Lambda)$ been used alone, the resulting
one-loop correction
which we denote with a bar symbol, would have been:
\eqn\zzx{ \bar Z_k^{(1)}(\Phi)={\lambda\Phi^2\over 192\pi^2}{2k^2+V''
\over{(k^2+V'')^2}}
={\lambda\Phi^2\over 192\pi^2}{2k^4+3k^2V''+{V''}^2\over{(k^2+V'')^3}},}
instead of \zza\ even though the limit $\bar Z^{(1)}_{k=0}(\Phi)
=\lambda^2\Phi^2/{192\pi^2 V''}$ is insensitive to whether
$\rho_k^{(4,1)}(s,\Lambda)$ is actually employed or not.
A comparison between \zzx\ and \zza\ reveals that the discrepancy between $\zk$
and $\bzk$ comes from an $O(k^4)$ mismatch in the numerator. The discrepancy
can be traced to be originated from terms that are multiplied by $p^2$ in
the derivative
expansion, i.e. the last two terms inside the curly bracket in \norw.
These are the quantities that vanish most
rapidly in the IR limit $k\to 0$ in the differential flow equation.
The insufficiency of using just $\rho_k^{(d)}(s,\Lambda)$
can be understood
as follows: From \spint, one readily sees that after the $p$ integration,
all $p^2$-dependent terms generated from derivative expansion
will acquire an extra factor of $s^{-1}$ compared with the ones without the
$p^2$ dependence. Subsequent $s$ integration using \rro\ then yields one
power of $(z^2+V'')$ for each $s^{-1}$. However, when $z$ is equated with $p$,
the cutoff result can no longer be recovered without the unjustified
substitution of $p^2$ by $p^2+V''$.
Therefore, to account for those higher order contributions that vanish
more rapidly as $k\to 0$, it is necessary to introduce
$\rho_k^{(d,1)}(s,\Lambda)$.

In the case where $d=4$, we have
\eqn\rtwy{\rho_k^{(4,1)}(s,\Lambda)=(1+k^2s+{1\over 2}k^4s^2)e^{-k^2s}
-(1+\Lambda^2s+{1\over 2}\Lambda^4s^2)e^{-\Lambda^2s},}
which differs from $\rho_k^{(4)}(s,\Lambda)$ by the higher-order
$s^2$-dependent
terms. These terms, as already demonstrated, are essential
for regaining the
expected cutoff dependence. That $\rho_k^{(d,1)}(s,\Lambda)$
provides a faster UV and IR
convergence can be seen from the fact that the smearing functions are of the
form of generalized incomplete $\Gamma$ function which can be expanded
as \ref\expan:
\eqn\ergd{ \rho_k^{(4,m)}(s,\Lambda)=e^{-k^2s}\sum_{\ell=0}^{m+1}
{1\over{\ell !}}(k^2s)^{\ell}-e^{-\Lambda^2s}\sum_{\ell=0}^{m+1}
{1\over{\ell !}}(\Lambda^2s)^{\ell}.}
This yields
\eqn\ferr{ \int_0^{\infty}ds~\rho^{(4,m)}_k(s,\Lambda)
e^{-\calh s}={1\over\calh}\Bigl[\Bigl({\Lambda^2\over{\calh+\Lambda^2}}
\Bigr)^{m+2}-\Bigl({k^2\over{\calh+k^2}}\Bigr)^{m+2}\Bigr],}
which indicates that the larger the $m$, the more rapid the convergence
\michael.
We also comment that choosing $m$ in
$\rho_k^{(4,m)}(s,\Lambda)$ is analogous to choosing
the number of ghost terms in the Pauli-Villars regularization since
in both approaches the divergence is eradicated by increasing the power of
$p$ dependence in the denominator. For example,
while only one ghost term is sufficient to regularize the logarithmic
divergence found in the four-point vertex function for the $\lambda\Phi^4$
theory in $d=4$, the quadratically divergent
integral characteristic of the two-point function calls for at least
two ghost terms to ensure the proper convergence \ref\cheng.

The requirement of using more than one smearing function to attain equality
between the momentum cutoff and operator cutoff regularizations
at the level of $\zk$ may seem disturbing at first glance since its
generalization to higher
orders of derivative expansion will become more complicated. However,
we remark that the computation of $\zk$ is generally dependent on how the
derivative terms are isolated; disagreements exist even within momentum
cutoff regularization itself. For example, imposing momentum
cutoff on the approach used by Fraser in \fraser\ would have lead to
\eqn\zfer{\tilde Z_k^{(1)}(\Phi)=
{\lambda\Phi^2\over 192\pi^2}{3k^2V''+{V''}^2\over{(k^2+V'')^3}},}
which also differs from \zza\ by a higher order $O(k^4)$ term in the
numerator. To reconcile the difference, we fist observe that since
the flow pattern of $\zk$ allows for the determination
of the anomalous dimension $\gamma_k$ via
\eqn\anoma{ \gamma_k=k{{\partial~{\rm ln}\zk}\over\partial k},}
the ambiguity in the $k$ dependence of $\zk$ must be intimately
connected with the scheme dependence in the computation of $\gamma_k$.
This is precisely what
was concluded in \ref\ball\ where the explicit cutoff dependence in
$\gamma_k$ beyond the lowest order was demonstrated. Such a dependence
should come as no
surprise since the RG coefficient functions such as the anomalous dimension
and $\beta$-functions are generally regularization dependent
beyond the leading order. Nonetheless, taking the physical limit $k\to 0$
for all three cases, we have
\eqn\ewer{ Z^{(1)}_{k=0}(\Phi)=\bar Z^{(1)}_{k=0}(\Phi)
=\tilde Z^{(1)}_{k=0}(\Phi)={1\over 192\pi^2}{\lambda^2\Phi^2\over V''}.}

Even though the one-loop correction to the wavefunction renormalization
exhibits different $k$ dependence for different methods, we argue that
when dealing with real physical situations the precise form
of the differential flow equations for the higher order derivative terms
should not be taken too seriously.
The concept of derivative expansion carried out in \lowsd\
has practical use only when one is interested in exploring the large-distance
effects in the IR regime. The expansion of $\tsk$ in powers
of $\parmu\Phi$ can only generate higher-order
corrections to the dominant
blocked potential $U_k(\Phi)$. In addition, we have seen that after
taking the physical limit $\Lambda\to\infty$ and
$k\to 0$, $U_{k=0}(\Phi)$ and $Z_{k=0}(\Phi)$ are all the same irrespective
of how they are computed. Hence, for all practical purpose
one may safely ignore the small higher order mismatches and
employ $\rhok$ alone to compute $U_k(\Phi)$, $\zk$ as well as other
higher order coefficients. While $\rhok$ can be regarded as
a sharp momentum cutoff for the leading order $U_k(\Phi)$, it corresponds
to a {\it smooth} momentum regulator for $\zk$ and beyond.
In the next Section where gauge theories are explored, we
shall see how these small mismatches precisely correspond to the
gauge noninvariant contributions that
must be purged in order to preserve gauge symmetry.

Another aspect concerning the use of a sharp cutoff in the derivative
expansion is the emergence of nonlocal interactions as we lower
the scale $k$ which defines the sharp boundary between the high and the
low modes.
The presence of nonlocality in $\tilde S_k[\Phi]$ is reflected by
the necessity of incorporating interactions to all ranges, and hence,
the simplifying picture of utilizing a reduced set of degrees of freedom
to characterize $\tilde S_k[\Phi]$ may be lost.
However, the objection against the use of sharp cutoff in conjunction
with derivative
expansion can be overcome by the following argument: In the perturbative
approach the loop integrations are performed between the IR and UV
cutoffs for summing up an infinite number of Feynman graphs for the partition
function. While the UV cutoff is eliminated by renormalization, we would
like to remove the IR cutoff as well in order to study the theory
in the thermodynamical limit $k\to 0$ where all physical observables
take on their limiting values. The goal of using a sharp IR cutoff
precisely allows us to explore these physical observables in the vicinity
$k \sim 0$.
Whatever nonlocality may arise from this sharp cutoff regularization will
also be present in the thermodynamical limit with any other regularization
schemes since the physics in this regime is independent of how one
achieves the elimination of all the high modes \ref\polonyi.
Similar viewpoints have also been presented by Morris in \ref\morris,
where the difficulties
and inadequacy of choosing a smooth regulator were
addressed. In fact the ``most promising'' method proposed there coincides
with the formalism we have developed earlier \lpp\ and presented here,
namely,
a derivative expansion around $k=0$ with a sharp cutoff as the candidate
for the low-energy effective theory.

\medskip
\bigskip

\centerline{\bf IV. OPERATOR CUTOFF AND GAUGE SYMMETRY}
\medskip
\nobreak
\xdef\secsym{4.}\global\meqno = 1
\medskip
\nobreak

In the previous sections we have seen that
by employing a set of
smearing functions $\rho_k^{(d,m)}(s,\Lambda)$ in the
operator cutoff regularization, the one-loop cutoff structure can be
recovered to arbitrary order in the derivative expansion. However,
the major distinction between operator cutoff and momentum cutoff
regularizations is that while the former is a special case of the proper-time
regulator and thus symmetry-preserving, the latter is not.
We now turn to gauge theories and explore the role dictated by symmetry.

In the course of evaluating the effective action for gauge theories,
one frequently encounters the following fluctuation operator:
\eqn\cdrw{ \calh=-D^2+\mu^2+Y(x),}
where $D_{\mu}$ is the covariant derivative for the gauge group, $\mu^2$
may be the mass for the scalar field coupled to the
gauge field
$A_{\mu}^a(x)$, and $Y(x)$
a matrix-valued function of $x$ describing, say the interaction
between the scalar particles. The index $a$ runs over the dimension of
the gauge group. One may also write $Y=Y^aT^a$ where $T^a$
are the generators of the gauge group satisfying ${\rm tr}(T^aT^b)
=-\delta^{ab}/2$ with ${\rm tr}$ denoting the summation over internal
indices only. When operating on $Y$, the covariant derivative gives
$D_{\mu}Y=\partial_{\mu} Y +[A_{\mu},Y]$ where $A_{\mu}=gA^a_{\mu}T^a$ with $g$
being a coupling constant. The unregularized one-loop contribution to
the effective action can be written as
\eqn\onfer{ \tilde S^{(1)}={1\over 2}{\rm Tr}\Bigl({\rm ln}\calh-{\rm ln}
\calh_0\Bigr)=-{1\over 2}\int_x\int_0^{\infty}{ds\over s}{\rm tr}\langle x|
\bigl(e^{-\calh s}-e^{-\calh_0s}\bigr)|x\rangle .}
The diagonal part of the ``heat kernel'' in the above can be written as
\eqn\diapl{\eqalign{ h(s;x,x)&=\langle x|e^{-\calh s}|x\rangle =\int_p
\langle x|p\rangle e^{-\calh_x s}\langle p|x\rangle
=\int_p e^{-ipx}e^{-\calh_xs}e^{ipx} \cr
&
=\int_pe^{-(p^2-2ip\cdot D+\calh_x)s}{\bf 1}
=\int_p e^{-(p^2+\mu^2)s}e^{(2ip\cdot D+D^2-Y)s}{\bf 1} \cr
&
=\sum_{n=0}^{\infty}{s^n\over n!}\int_p e^{-(p^2+\mu^2)s}\bigl(2ip\cdot D+D^2-Y
\bigr)^n{\bf 1},}}
where we have employed the plane wave basis $|p\rangle$ with
$\langle x|p\rangle=e^{-ipx}$ and the following commutation relations
\chiral, \ref\nepomechie:
\eqn\rela{ [D_{\mu}, e^{ipx}]=ip_{\mu}, \qquad\qquad
[\calh_x,e^{ipx}]=p^2-2ip\cdot D.}
The factor ${\bf 1}$ indicates that the operator $D_{\mu}$ acts on the
identity.
Inserting \diapl\ into \onfer, we are lead to
\eqn\ferg{\eqalign{ \tilde S^{(1)} &=-{1\over 2}\sum_{n=0}^{\infty}{1
\over n!}\int_x\int_0^{\infty}ds~s^{n-1}\int_p {\rm tr}\Biggl\{
e^{-(p^2+\mu^2)s}\bigl(2ip\cdot D+D^2-Y\bigr)^n\Biggr\}{\bf 1} \cr
&
={1\over 2}\int_x\int_p{\rm tr}\Biggl\{{\rm ln}\bigl(p^2+\mu^2\bigr)
-\sum_{n=1}^{\infty}{1\over n}\Bigl(
{{2ip\cdot D+D^2-Y}\over{p^2+\mu^2}}\Bigr)^n\Biggr\}{\bf 1},}}
where the first term in the integrand represents an overall constant
and can be subsequently dropped.
The above expression suggests that it is possible to expand the effective
action in terms of inverse mass provided that the background fields
$A_{\mu}(x)$ and $Y(x)$ vary slowly on the scale $\mu^{-1}$. Thus, we
may write the effective action as \ref\dyakonov-\ref\bilenky:
\eqn\cdfa{ \tilde S^{(1)}=\int_x\sum_{n=1}^{\infty}{a_n\over {(\mu^2)}^{n-2}}
\calo_n,}
where $\calo_n$ are the traces of dimension $2n$ gauge invariant operators
and can in general be written as
\eqn\cfter{ \calo_n=\sum_{i}\gamma_i^{(n)}\tilde\calo_n^{(i)},}
with $\gamma_i^{(n)}$ being the numerical coefficients associated
with $\tilde\calo_n^{(i)}$, the set of linearly independent traces.
As an example, using the field strength $F_{\mu\nu}$,
$Y$ and $D_{\mu}$ one can construct the following:
\eqn\gauinv{\eqalign{\tilde\calo_1 &=\bigl\{ {\rm tr}(Y)\bigr\}, \cr
\tilde\calo_2 &=\bigl\{{\rm tr}(Y^2),~{\rm tr}(F_{\mu\nu}F_{\mu\nu})\bigr\},\cr
\tilde\calo_3&=\bigl\{{\rm tr}(Y^3),~{\rm tr}({D_{\mu}Y})^2,~
{\rm tr}(F_{\mu\nu}YF_{\mu\nu}),~{\rm tr}(D_{\mu}F_{\mu\nu}D_{\sigma}
F_{\sigma\nu}),~{\rm tr}(F_{\mu\nu}F_{\nu\sigma}F_{\sigma\mu})\bigr\}.}}

With the integrand written explicitly in terms of symmetry-preserving
quantities,
gauge invariance of the effective action is automatically fulfilled by
choosing an invariant regularization scheme. One possible candidate
for regulating the second expression on the right-hand-side of \ferg\
is by dimensional regularization since the non-invariant
momentum cutoff is undesirable here.
However, if we consider the first equation in \ferg\ instead,
the most natural way to do away the divergence is to introduce a set of
smearing functions $\rho^{(d,m)}_k(s,\Lambda)$ for the proper-time integration,
as suggested before.
The advantage of going to the proper-time formalism is that in
addition to allowing the
theory to be regulated in a completely
invariant manner by preserving the full symmetries, even gauge symmetry,
of the original lagrangian, it also admits cutoff scales.
The symmetry-preserving nature of the regulator can easily be seen
from the absence of cutoff scales in the $p$ integration and the transfer of
spacetime singularity into a singularity in the proper-time variable
which is independent of symmetry transformation on the background fields.
The insertion of the regulating function may be thought of as solely
for the purpose
of ``technical'' convenience to cope with the divergence manifested
in the $s$ integration. Any possible violation of
symmetry within the operator cutoff formalism can take place only if
the smearing function depends on certain parameters
such as the background fields or $p$. In such case, the regularized action
will vary under symmetry transformation.

Below we regularize the theory with both the gauge
non-invariant momentum cutoff and the invariant operator cutoff
regulators to establish a connection between them.
\medskip
\medskip
\centerline{\bf a. momentum cutoff regularization}
\medskip

The one-loop contribution to the effective blocked action regularized
by momentum cutoff can be written as
\eqn\ferg{\eqalign{ \tilde S_k^{(1)} &=
{1\over 2}\int_x\int_p^{'}{\rm tr}\Bigl\{{\rm ln}
\bigl(p^2+\mu^2-2ip\cdot D-D^2+Y\bigr)\Bigr\}{\bf 1} \cr
&
= -{1\over 2}\int_x\int_0^{\infty}{ds\over s}\int_p^{'} {\rm tr}\Bigl\{
e^{-(p^2+\mu^2)s}e^{(2ip\cdot D+D^2-Y)s}\Bigr\}{\bf 1}.}}
Following the details presented by Nepomechie \nepomechie\ and Mukku
\ref\mukku, we first employ the Baker-Campbell-Hausdorf formulae:
\eqn\bch{ e^{A+B}=e^A\Biggl\{1+B+{[B,A]\over 2}+{B^2\over 2}
+{{[[B,A]+B^2,A]}\over 3!}+{[B,A]B\over 3!}+{B^3\over 3!}
+\cdots\Biggr\},}
\eqn\bchh{\eqalign{e^{A+B+C}&=e^{A+B}\Biggl\{1+C+{[C,A]\over 2}
+{[C,B]\over 2}+{C^2\over 2}+{[[C,A],A]\over 3!}+{[[C,A],B]
\over 3!} +{[[C,B],A]\over 3!} \cr
&
+{[[C,B],B]\over 3!}+{[C,[C,A]]\over 3!}
+{[C,[C,B]]\over 3!}+{[C,A]C\over 2}+{[C,B]C\over 2}+{C^3\over 3!}
+\cdots\Biggr\},}}
and expand the heat kernel as
\eqn\egrt{\eqalign{ h(s;x,x)&=\int_p^{'} e^{-(p^2+\mu^2)s}e^{(2ip\cdot
D+D^2-Y)s}~{\bf 1} \cr
&
=e^{-(\mu^2+Y)s}\int_p^{'}e^{-p^2s}
\Biggl\{1+D^2s-2p_{\mu}p_{\nu}D_{\mu}D_{\nu}s^2+{D^4s^2\over 2}
-{[D^2,Y]s^2\over 2} \cr
&
-{2\pmu\pnu \over 3}\Bigl\{ [[D^2,\dmu],\dnu]+3\dnu[D^2,\dmu]
+3\dmu\dnu D^2-[\dmu\dnu,Y] \cr
&
-[\dmu,Y]\dnu
-p_{\alpha}p_{\beta}\dmu\dnu D_{\alpha}
D_{\beta}s\Bigr\}s^3+\cdots\Biggr\}~{\bf 1},}}
where the contributions with odd powers of $p$ are neglected since they
give vanishing contribution
after momentum integrations. Substituting the above into \ferg\ and
carrying out the $s$ integral, we obtain
\eqn\efs{\eqalign{\tilde S_k^{(1)}&={1\over 2}\int_x\int_p^{'}{\rm tr}~
\Biggl\{{\rm ln}\bigl(p^2+\mu^2+Y\bigr)-{1\over{p^2+\mu^2+Y}}D^2
+{2\pmu\pnu\over{(p^2+\mu^2+Y)^2}}\dmu\dnu \cr
&
-{1\over 2(p^2+\mu^2+Y)^2}\bigl(D^4-[D^2,Y]\bigr)
+{4\pmu\pnu\over 3(p^2+\mu^2+Y)^3}\biggl([[D^2,\dmu],\dnu]+3\dnu[D^2,\dmu]\cr
&
+3\dmu\dnu D^2
-[\dmu\dnu,Y]-[\dmu,Y]\dnu\biggr)
-{4\pmu\pnu{p_{\alpha}}p_{\beta}\over{(p^2+\mu^2+Y)^4}}\dmu\dnu D_{\alpha}
D_{\beta}+\cdots\Biggr\}~{\bf 1},}}
which by rotational $O(d)$ symmetry can be further simplified to
\eqn\efso{\eqalign{\tilde S_k^{(1)}&={1\over 2}\int_x\int_p^{'}{\rm tr}~
\Biggl\{{\rm ln}\bigl(p^2+\mu^2+Y\bigr)+{1\over{(p^2+\mu^2+Y)^2}}\Bigl[
{(2-d)p^2-d(\mu^2+Y)\over d}D^2 \cr
&
-{1\over 2}\bigl(D^4-[D^2,Y]\bigr)\Bigr]
+{4p^2\over 3d(p^2+\mu^2+Y)^3}\Bigl([[D^2,\dmu],\dmu]+3\dmu[D^2,\dmu]
+3D^4 \cr
&
-[D^2,Y]-[\dmu,Y]\dmu\Bigr)
-{4{(p^2)}^2\over{d(d+2)(p^2+\mu^2+Y)^4}}\Bigl[D^4+(\dmu\dnu)^2+
\dmu D^2\dmu\Bigr]\Biggr\}~{\bf 1},}}
and evaluated with \ods. Finally, using
\eqn\ope{\dmu Y=(\dmu Y)+Y\dmu,\qquad D^2Y=(D^2Y)+2\dmu Y\dmu-YD^2,
\qquad F_{\mu\nu}=[\dmu,\dnu],}
eq.\efso\ becomes
\eqn\efsor{\eqalign{\tilde S_k^{(1)}&={1\over 2}\int_x\int_p^{'}{\rm tr}~
\Biggl\{ {\rm ln}\bigl(p^2+\mu^2+Y\bigr)+a_1D^2+a_2(D^2Y)+a_3\dmu Y\dmu
+a_4YD^2 \cr
&\qquad\qquad\qquad
+a_5D^4+a_6\dmu D^2\dmu+a_7F_{\mu\nu}F_{\mu\nu}\Biggr\}~{\bf 1},}}
where
\eqn\aone{ a_1={{(2-d)p^2-d(\mu^2+Y)}\over{d(p^2+\mu^2+Y)^2}},}
\eqn\atwo{a_2={{(3d-8)p^2+3d(\mu^2+Y)}\over{6d(p^2+\mu^2+Y)^3}},}
\eqn\athree{a_3=-a_4={{(d-4)p^2+d(\mu^2+Y)}\over{d(p^2+\mu^2+Y)^3}},}
\eqn\afive{a_5=-{{(3d-2)(d-4)p^4+2p^2(d+2)(3d-8)(\mu^2+Y)+3d(d+2)(\mu^2+Y)^2}
\over{6d(d+2)(p^2+\mu^2+Y)^4}},}
\eqn\asix{a_6={{4p^2\bigl[(d-4)p^2+(d+2)(\mu^2+Y)\bigr]}
\over{3d(d+2)(p^2+\mu^2+Y)^4}},}
and
\eqn\aseven{ a_7=-{2{(p^2)}^2\over{d(d+2)(p^2+\mu^2+Y)^4}}.}
The additional noninvariant operators
$D^2$, $\dmu Y\dmu$, $YD^2$, $D^4$ and $\dmu D^2\dmu$
generated by the momentum cutoff regulator in \efsor\ can be readily seen
from a simple comparison with \cdfa\ which is consisted of gauge invariant
quantities only. Nevertheless, taking the limit
$\Lambda\to\infty$ and $k\to 0$, the coefficients associated with these gauge
noninvariant contributions are identically zero, i.e.
\eqn\nogau{ \int_pa_i=0,}
for i=1, 3, 4, 5 and 6. The presence of noninvariant operators for
theories regularized by momentum cutoff makes it difficult to extend
the regularization to gauge theories. One must therefore resort to other
methods which contain the cutoff scales and yet preserve the symmetries
of the original lagrangian. A promising candidate which encompasses both
features is the operator cutoff regularization which we next turn to.

\medskip
\medskip
\centerline{\bf b. operator cutoff regularization}
\medskip

We now apply the alternative operator cutoff formalism to regularize
the divergence found in \onfer. Following the methodology outlined in
the previous section, the heat kernel \egrt\ may modified as
\eqn\egrt{\eqalign{ h(s;x,x)&\to h(s;x,x)\Big\vert_{\rm oc}
=e^{-(\mu^2+Y)s}\int_pe^{-p^2s}\Biggl\{\rho^{(d,0)}_k(s,\Lambda)
\bigl(1+D^2s+{D^4\over 2}s^2-{[D^2,Y]\over 2}s^2\bigr) \cr
&
-{2p^2\over d}\rho^{(d,1)}_k(s,\Lambda)\Bigl[D^2s^2+{1\over 3}
\Bigl([[D^2,\dmu],\dmu]+3\dmu[D^2,\dmu]
+3D^4 -[D^2,Y] \cr
&
-[\dmu,Y]\dmu\Bigr)s^3\Bigr]
+{2{(p^2)}^2\over 3d(d+2)}\rho^{(d,2)}_k(s,\Lambda)\Bigl[D^4
+(\dmu\dnu)^2+\dmu D^2\dmu\Bigr]s^4\Biggr\}~{\bf 1} \cr
&
={e^{-(\mu^2+Y)s}\over (4\pi s)^{d/2}}\Biggl\{ \rho_k^{(d,0)}(s,\Lambda)
+b_1D^2+b_2(D^2Y)+b_3\dmu Y\dmu+b_4YD^2+b_5D^4\cr
&\qquad\qquad
+b_6\dmu D^2\dmu+b_7F_{\mu\nu}F_{\mu\nu}+O(s^3)\Biggr\}~{\bf 1},}}
where, to have agreements with the cutoff results, one requires
\eqn\bone{b_1=\bigl[\rho_k^{(d,0)}(s,\Lambda)-\rho_k^{(d,1)}(s,\Lambda)\bigr]s
={2s^{1+d/2}\over{S_d\Gamma(d/2)}}\int_z^{'}\bigl(1-{2\over d}z^2s\bigr)
e^{-z^2s},}
\eqn\btwo{b_2=-{1\over 6}\bigl[3\rho_k^{(d,0)}(s,\Lambda)-2\rho_k^{(d,1)}
(s,\Lambda)\bigr]s^2=-{s^{2+d/2}\over{S_d\Gamma(d/2)}}\int_z^{'}\bigl(1
-{4\over 3d}z^2s\bigr)e^{-z^2s},}
\eqn\bthree{b_3=-b_4=-\bigl[\rho_k^{(d,0)}(s,\Lambda)-\rho_k^{(d,1)}
(s,\Lambda)\bigr]s^2=-{2s^{2+d/2}\over{S_d\Gamma(d/2)}}\int_z^{'}
\bigl(1-{2\over d}z^2s\bigr)e^{-z^2s},}
\eqn\bfive{\eqalign{b_5&={1\over 6}\bigl[3\rho_k^{(d,0)}(s,\Lambda)
-4\rho_k^{(d,1)}
(s,\Lambda)+\rho_k^{(d,2)}(s,\Lambda)\bigr]s^2 \cr
&
={s^{2+d/2}\over{3S_d\Gamma(d/2)}}\int_z^{'}
\Bigl(3-{8\over d}z^2s+{4\over{d(d+2)}}{(z^2s)}^2\Bigr)e^{-z^2s},}}
\eqn\bsix{b_6=-{1\over 3}\bigl[\rho_k^{(d,1)}(s,\Lambda)-\rho_k^{(d,2)}
(s,\Lambda)\bigr]s^2=-{4s^{2+d/2}\over{3dS_d\Gamma(d/2)}}\int_z^{'}
\bigl(1-{2\over {d+2}}z^2s\bigr)z^2se^{-z^2s},}
and
\eqn\bseven{b_7={1\over 12}\rho_k^{(d,2)}(s,\Lambda)s^2
={s^{2+d/2}\over{6S_d\Gamma(d/2)}}\int_z^{'}{(z^2s)}^2e^{-z^2s}.}
Notice that there exists a one-to-one correspondence between
the coefficients $b_i's$ in the proper-time formulation and the
$a_i's$ in the momentum cutoff approach.
In arriving at \egrt, we again have inserted the regulating smearing
functions $\rho_k^{(d,m)} (s,\Lambda)$ whose general form has been derived
in \tyre, and performed the momentum integrations using
\eqn\cvns{ \int_pp_{\mu_1}p_{\mu_2}\cdots p_{\mu_{2m}}e^{-p^2s}
={{T^m_{\mu_1\mu_2\cdots\mu_{2m}}\Gamma(d/2)}\over{2^m\Gamma(m+d/2)}}
\int_p{(p^2)}^me^{-p^2s}
={{T^m_{\mu_1\mu_2\cdots\mu_{2m}}}\over{(4\pi s)^{d/2}(2s)^m}}.}
However, our aim here is not to reproduce the sharp cutoff results, but to
formulate a regularization scheme which leads to a gauge invariant blocked
action $\tilde S_k$. Gauge symmetry remains unbroken only if the
contributions from $b_i$ for i=1, 3, 4, 5 and 6 vanish. This requirement
is easily attained if instead of using $\rho_k^{(d,m)}(s,\Lambda)$, only
$\rho_k^{(d,0)}(s,\Lambda)=\rhok$ is utilized in the expansion of the heat
kernel
in \egrt. The resulting gauge invariant ``blocked'' heat kernel then
takes on the form
\eqn\egrst{ h_k(s;x,x)={e^{-(\mu^2+Y)s}\over
(4\pi)^{d/2}}\rho_k^{(d)}(s,\Lambda)\Biggl\{
1+{1\over 12}\Bigl[F_{\mu\nu}F_{\mu\nu}-2(D^2Y)\Bigr]s^2+O(s^3)\Biggr\},}
which, for $\rho_k^{(d)}(s,\Lambda)\to 1$, agrees with that obtained in
\nepomechie\ and \mukku. The one-loop blocked action becomes
\eqn\blop{\eqalign{\tilde S^{(1)}_k &=-{1\over 2(4\pi)^{d/2}}\int_x
\int_0^{\infty}{ds\over s^{1+d/2}}\rho_k^{(d)}(s,\Lambda)e^{-(\mu^2+Y)s}
\Biggl\{1+{1\over 12}
\Bigl[F_{\mu\nu}F_{\mu\nu}-2(D^2Y)\Bigr]s^2\Biggr\}\cr
&
=-{1\over 2}\int_x\int_0^{\infty}{ds\over s}e^{-(\mu^2+Y)s}
\Biggl\{1+{1\over 12}
\Bigl[F_{\mu\nu}F_{\mu\nu}-2(D^2Y)\Bigr]s^2\Biggr\}\int_z^{'}e^{-z^2s}.}}
The RG flow equation can subsequently be obtained by varying \blop\ with
respect to the IR scale $k$.

What we have seen here is that by using just one smearing function
$\rho_k^{(d)}(s,\Lambda)$, an effective
blocked action having a momentum cutoff regularized
scalar sector as well as the symmetry-preserving contributions from
the gauge fields is obtained. The invariant prescription adopted here
departs from the usual momentum cutoff regularization in the sense
that the gauge noninvariant contributions are effectively
subtracted off. Any dependence on the UV cutoff $\Lambda$ present in
\blop\ can subsequently be absorbed with the usual procedure of
renormalization, i.e. redefinition of parameters. In particular, the
familiar ${\rm ln}\Lambda^2$ divergence coming from the
$F_{\mu\nu}F_{\mu\nu}$ term
for $d=4$ can be dialed away via the coupling constant renormalization.
{}From the modification induced by $\rho_k^{(d)}(s,\Lambda)$ on
the fluctuation operator $\calh$:
\eqn\hhm{ \calh\rightarrow\calh\Big\vert_{\rm oc}
=\calh\Bigl(1+{\calh\over\Lambda^2}
\Bigr)^2=\calh +{2\calh^2\over\Lambda^2}+\cdots,}
one readily notices the
similarity between the effect brought about by the operator cutoff
regularization and the gauge invariant
method of higher covariant derivatives \ref\faddeev.

We conclude this section with the remark that the task of preserving
gauge symmetry for any given
regularization often amounts to finding a proper way of
transferring the singularity accompanied in the trace operation to
some parameters which are independent of gauge transformation. For example,
the spirit of dimensional regularization is to displace $d$, the
space dimensionality in which the system is defined, by a small
positive quantity
$\epsilon$. Since gauge symmetry is not influenced by the value of $d$,
gauge invariance
is readily fulfilled by transforming the divergent
structures of the theory into a pole term $\sim\epsilon^{-1}$. In a
similar fashion, operator cutoff approach transfers the divergences to
the proper-time
parameter $s$. Embedding the cutoff scales tactically in the
regulating smearing function $\rho_k^{(d)}(s,\Lambda)$ leads to an
effective blocked action which is manifestly gauge invariant.
That the momentum cutoff regularization fails to be an invariant prescription
is seen here from its requirement of having to employ more than one
smearing function when expressed in the proper-time
representation, and hence it does not fall into the generalized
class of invariant proper-time regularization.

\bigskip
\medskip
\centerline{\bf V. SUMMARY AND DISCUSSIONS}
\medskip
\nobreak
\xdef\secsym{5.}\global\meqno = 1
\medskip

\nobreak

In this paper we have followed the formalism developed in ref. \michael\
and illustrated how a theory regularized with momentum cutoff can be
represented by the proper-time parameterization using a set of regulating
smearing functions $\rho_k^{(d,m)}(s,\Lambda)$.
These smearing functions incorporate cutoff scales and reproduce
the essential features of blocking transformation, albeit in the less obvious
proper-time coordinate.
The modifications induced by $\rhok$ in our operator
cutoff regularization
are seen to be reminiscent to that of the Pauli-Villars, or the
generalized method of higher (covariant) derivatives.

Equivalence between the two regulators
was demonstrated for the one-loop effective blocked action expanded in
number of (covariant) derivatives.
We also computed the one-loop corrections to the
two- and four-point functions for the $\lambda\Phi^4$ theory and showed
that the cutoff expressions can indeed be reproduced order by
order in terms of coupling constant $\lambda$ using the operator cutoff
formalism.

The most important feature of the operator cutoff regularization
is that it is a gauge invariant prescription even when momentum cutoff
scales are
present. The symmetry-preserving nature of the formalism
is attributed to its capacity of transferring the singularity that arises
from taking the spacetime
trace to the proper-time variable $s$ which is independent
of gauge transformation. Gauge invariance is ensured by retaining
the $s$ integration to be performed last.

Instead of employing the entire set of smearing functions $\rho_k^{(d,m)}
(s,\Lambda)$, the invariant prescription we proposed here is to
utilize only $\rhok=\rho_k^{(d,0)}(s,\Lambda)$.
While the momentum cutoff structure for the leading order blocked potential is
automatically reproduced with $\rhok$ alone, discrepancies
between the two formalisms inevitably arise in the high order (covariant)
derivative terms. The differences, as noted in Sec. IV,
are precisely the gauge noninvariant contributions that are generated
in the momentum cutoff prescription.
The virtue of $\rhok$ is that it is chosen in a such a way
that the gauge noninvariant sector in the effective theory
are completely relegated. Therefore, our invariant regularization resembles
a sharp cutoff for the blocked potential and a smooth regulator for the
derivative terms. The requirement of using the complete
set of $\rho_k^{(d,m)}(s,\Lambda)$ in order to reproduce the cutoff results
term by term in the derivative expansion provides another indication
that momentum cutoff does not belong to the generalized class of proper-time
regularization and hence cannot be a gauge invariant prescription.

With momentum scales implemented in a symmetry-preserving manner,
operator cutoff regularization offers
a promising method for exploring the RG flow
of gauge theories in the spirit of Wilson-Kadanoff blocking transformation.
The evolution of the theory will now be characterized by the variation
of the blocked action in response to the change in the proper-time
smearing function $\rhok$. For example, the full non-linear
RG flow equation for the
scalar theory explored in Sec. II can be written as
\eqn\fll{ k{{\partial U_k(\Phi)}\over{\partial k}}
=-{1\over 2(4\pi)^{d/2}}\int_0^{\infty}{ds\over s^{1+d/2}}
\bigl(\krk\bigr)\Bigl(e^{-U_k''(\Phi)s}-e^{-U_k''(0)s}\Bigr).}
The caution to be taken when employing operator cutoff regularization
for Yang-Mills theory is that BRS symmetry is generally violated unless
a covariant background
field gauge is chosen \ref\rebhan. The general framework of the covariant
technique for computing the one-loop effective action can be found
in \ref\avramidi.
In a preliminary work \ref\ym, we adopted the spirit of
operator cutoff outlined above and demonstrated
how the expected $\beta$ function and the corresponding
RG flow of the Yang-Mills theories can be obtained with
the Wilson-Kadanoff blocking approach. A more thorough study for
the non-abelian theories is under way.

\medskip
\medskip
\centerline{\bf ACKNOWLEDGEMENTS}
\medskip
\nobreak
The author is grateful to K. Johnson, B. M\"uller, M. Oleszczuk and
J. Polonyi for
careful reading of the manuscript and valuable comments.

\goodbreak
\bigskip
\medskip
\goodbreak
\centerline{\bf REFERENCES}
\medskip
\nobreak
\medskip

\par\hang\noindent{\thooft} G. 't Hooft and M. Veltman, {\it Nucl. Phys.}
{\bf B44} (1971) 189.
\medskip
\par\hang\noindent{\dowker} J. S. Dowker and R. Critchley, {\it Phys. Rev.}
{\bf D13} (1976) 3224.
\medskip
\par\hang\noindent{\schwinger} J. Schwinger, {\it Phys. Rev.} {\bf 82}
(1951) 664.
\medskip
\par\hang\noindent{\pauli} W. Pauli and F. Villars, {\it Rev. Mod. Phys.}
{\bf 21} (1949) 434.
\medskip
\par\hang\noindent{\wilson} K. Wilson and J. Kogut, {\it Phys. Rep.}
{\bf 12C} (1975) 75; L. Kadanoff, {\it Physics} {\bf 2} (1966) 263.
\medskip
\par\hang\noindent{\lp} S.-B. Liao and J. Polonyi, {\it Ann. Phys.}
{\bf 222} (1993) 122.
\medskip
\par\hang\noindent{\wetterich} A. Ringwald and C. Wetterich, {\it Nucl.Phys.}
{\bf B334} (1990) 506; C. Wetterich, {\it Nuc. Phys.} {\bf B352}
(1990) 529.
\medskip
\par\hang\noindent{\finite} S.-B. Liao, J. Polonyi and D.-P. Xu,
{\it Phys. Rev.} {\bf D51} (1995) 748; S.-B. Liao
and J. Polonyi, {\it Nucl. Phys.} {\bf A570} (1994) 203c.
\medskip
\par\hang\noindent{\ma} see for example, S.-K. Ma, {\it Modern Theory
of Critical Phenomena} (Benjamin, Reading, 1976), p. 530.
\medskip
\par\hang\noindent{\wegner} F.J. Wegner and A. Houghton, {\it Phys. Rev.}
{\bf A8} (1972) 401.
\medskip
\par\hang\noindent{\cwetterich} M. Reuter and C. Wetterich,
{\it Phys. Lett.} {\bf B334} (1994) 412;
U. Ellwanger, C. Wetterich, {\it Nucl. Phys.} {\bf B423} (1994) 137.
\medskip
\par\hang\noindent{\michael} M. Oleszczuk, {\it Z. Phys.} {\bf C64}
(1994) 533.
\medskip
\par\hang\noindent{\chiral} R. D. Ball, {\it Phys. Rep.} {\bf 182} (1989) 1.
\medskip
\par\hang\noindent{\zuk} J. Zuk, {\it Int. J. Mod. Phys.} {\bf A18} (1990)
3549.
\medskip
\par\hang\noindent{\lpp} S.-B. Liao and J. Polonyi,
{\it Phys. Rev.} {\bf D51} (1995) 4474.
\medskip
\par\hang\noindent{\ft} S.-B. Liao and M. Strickland, {\it Phys. Rev.} {\bf
D52} (1995) 3653.
\medskip
\par\hang\noindent{\grad} J. A. Zuk, {\it Nucl. Phys.} {\bf B280} (1987)
125, {\it Phys. Rev.} {\bf D34} (1986) 1791 and {\bf D32} (1985) 2653;
L-H. Chan, {\it Phys. Rev. Lett.} 55 (1985) 1222; J. Iliopoulos,
C. Itzykson and A. Martin, {\it Rev. Mod. Phys.} {\bf 47} (1975) 165.
\medskip
\par\hang\noindent{\fraser} C. M. Fraser, {\it Z. Phys.} {\bf C28} (1985) 101;
I. Aitchison and C. Fraser, {\it Phys. Rev.} {\bf D32} (1985) 2190.
\medskip
\par\hang\noindent{\expan} L. C. Andrews, {\it Special functions of
mathematics for engineers}, (McGraw Hills, New York, 1992), p. 88.
\medskip
\par\hang\noindent{\cheng} T. Cheng and L. Li, {\it Gauge theory of elementary
particle physics}, (Clarendon, Oxford, 1984), p. 50.
\medskip
\par\hang\noindent{\ball} R. D. Ball, P. E. Haggensen, J. I. Latorre and
E. Moreno, {\it Phys. Lett.} {\bf B347} (1995) 80.
\medskip
\par\hang\noindent{\polonyi} We thank Professor J. Polonyi for discussions
on this point.
\medskip
\par\hang\noindent{\morris} T. R. Morris, {\it Int. J. Mod. Phys.} {\bf A9}
(1994) 2411; SHEP-95-21/hep-th/9508017; {\it Phys.Lett.} {\bf B334}
(1994) 355 and {\bf B329} (1994) 241.
\medskip
\par\hang\noindent{\nepomechie} R. I. Nepomechie, {\it Phys. Rev.} {\bf D31}
(1985) 3291.
\medskip
\par\hang\noindent{\dyakonov} D. I. D'yakonov, V. Yu. Petrov and A. V. Yung,
{\it Sov. J. Nucl. Phys.} {\bf 39} (1984) 150.
\medskip
\par\hang\noindent{\bilenky} M. Bilenky and A. Santamaria, {\it Nucl.
Phys.} {\bf B420} (1994) 47.
\medskip
\par\hang\noindent{\mukku} C. Mukku, {\it Phys. Rev.}
{\bf D45} (1992) 2916.
\medskip
\par\hang\noindent{\faddeev} L. D. Faddeev and A. A. Slavnov, {\it Gauge
fields} 2nd ed. (Benjamin, Reading, 1990).
\medskip
\par\hang\noindent{\rebhan} A. Rebhan, {\it Phys. Rev.} {\bf D39} (1989) 3101.
\medskip
\par\hang\noindent{\avramidi} see for example, B. S. De Witt, {\it Phys. Rep.}
{\bf 19} (1975) 296; I. G. Avramidi, {\it Phys. Lett.} {\bf B238}
 (1990) 92 and {\it Nucl. Phys.} {\bf B355} (1991) 712.
\medskip
\par\hang\noindent{\ym} S.-B. Liao, {\it Chin. J. Phys.} {\bf 32} (1994) 1109;
and {\it Proceeding of the Workshop on Quantum Infrared Physics}, ed. by
H. M. Fried and B. M\"uller, (World Scientific, Singapore, 1995), p. 83.
\medskip
%
\end